# Observation of Chromospheric Sunspot at Millimeter Range with the Nobeyama 45 m Telescope

Kazumasa Iwai[1] and Masumi Shimojo[2]

1, Nobeyama Solar Radio Observatory, National Astronomical Observatory of Japan, Minamimaki, Nagano 384-1305, Japan; kazumasa.iwai@nao.ac.jp
2, National Astronomical Observatory of Japan, Mitaka, Tokyo 181-8588, Japan

## ABSTRACT

The brightness temperature of the radio free-free emission at millimeter range is an effective tool for characterizing the vertical structure of the solar chromosphere. In this paper, we report on the first single-dish observation of a sunspot at 85 and 115 GHz with sufficient spatial resolution for resolving the sunspot umbra using the Nobeyama 45 m telescope. We used radio attenuation material, i.e. a solar filter, to prevent the saturation of the receivers. Considering the contamination from the plage by the side-lobes, we found that the brightness temperature of the umbra should be lower than that of the quiet region. This result is inconsistent with the preexisting atmospheric models. We also found that the brightness temperature distribution at millimeter range strongly corresponds to the ultraviolet (UV) continuum emission at 1700 Å, especially at the quiet region.

## 1. INTRODUCTION

Atmospheric models of sunspots have been developed from optical and ultraviolet (UV) line observations (e.g. Avrett 1981). Since these lines are formed under non-local thermodynamic equilibrium (non-LTE) conditions, line formation regions are not well understood, and radiative transfer simulations are required to derive the physical conditions of their emission region.

The main emission mechanism of the Sun in millimeter and submillimeter wavelength range is a thermal free-free emission from the chromosphere. This emission is formed under LTE conditions. The opacity of the thermal free-free emission is determined by the electron temperature and density (Dulk 1985). In addition, the Rayleigh-Jeans law can be applied in this wavelength range. Hence, the observed brightness temperature can be converted into the thermal electron temperature of the radio source region. However, the radio source region of the free-free emission of the Sun at a given wavelength is distributed over a significant range of heights. For example, the emission height at 3 mm range is widely distributed between the chromosphere and the transition region (Wedemeyer-Böhm et al. 2007). Hence, wideband observations that cover the submillimeter, millimeter, and centimeter range are required to derive an accurate model of the solar atmosphere.

Large millimeter and submillimeter telescopes usually cannot observe the Sun because they are not designed to observe such a high-brightness body. Hence, there have been only a few sunspot observations at this wavelength range. At the submillimeter wavelength range, the James Clerk Maxwell Telescope (JCMT) and the Caltech Submillimeter Observatory (CSO) have been used for this purpose (Lindsey & Kopp 1995; Lindsey et al. 1995; Bastian et al. 1993). These studies suggested that the brightness temperature of a sunspot umbra is lower than that of the quiet region.

At millimeter range, a larger diameter single-dish observation or interferometer observation are required to resolve a sunspot umbra because of the longer wavelength than the submillimeter range. For example, White et al. (2006) observed sunspots using an interferometer (Berkeley-Illinois-Maryland Array [BIMA]) at 85 GHz. From this observational result, Loukitcheva et al. (2014) suggested that the upper limit of the brightness temperature of the sunspot umbra was almost the same as, or lower than, that of the quiet region. Millimetric interferometers, however, usually have limited baselines. Hence, they have different sensitivities depending on the radio source sizes. They are therefore unsuitable for observing an absolute brightness temperature of a broad radio source such as a solar sunspot. On the other hand, a single-dish telescope with a diameter large enough to resolve the sunspot structures is effective for sunspot observation. The Nobeyama 45 m radio telescope, for example, is a single-dish telescope used for millimetric radio astronomy. The typical spatial resolution of this telescope is about 15" at 115 GHz. Thus it enables us to successfully distinguish the sunspot umbra from the surrounding regions.

Having discussed these practical considerations that bear upon sunspot observation, it is important to note that there are many atmospheric models for sunspots. In particular, Loukitcheva et al. (2014) calculated the radio brightness temperature of sunspots from various atmospheric models. They showed that many atmospheric models suggest that sunspots are darker than the quiet region at submillimeter range, which is consistent with the results suggested by the submillimeter observations (e.g. Lindsey & Kopp 1995). At millimeter range, on the other hand, most of the models suggest that sunspots are brighter than the quiet region.

In this study, we measure the relative brightness of various chromospheric features at millimeter range by using the Nobeyama 45 m telescope, which marks the first single-dish observation of a sunspot at millimeter range with sufficient spatial resolution for resolving the sunspot umbra. We then compare these results against the predictions of various atmospheric models. The instrument used in this study is described in Section 2. The observational results and data analysis are provided in Section 3 and discussed in Section 4. Lastly, we summarize this study in Section 5.

## 2. INSTRUMENTS
The Nobeyama 45 m radio telescope is a ground-based, millimeter-range radio telescope operated by the Nobeyama Radio Observatory (NRO), National Astronomical Observatory of Japan (NAOJ). This telescope has been used to observe large solar structures, such as the polar brightening or prominences (Kosugi et al. 1986; Irimajiri et al. 1995). We used two Superconductor-Insulator-Superconductor (SIS) receivers (S80 and S100). These two receivers can be tuned at different frequencies and can receive radio emissions simultaneously. We observed at 85 GHz (3.5 mm) and 115 GHz (2.6 mm). The typical spatial resolutions of this telescope were 19" and 15" at 85 and 115 GHz, respectively.

We used a radio attenuation material – a so-called "solar filter" – to prevent the receiver system from becoming saturated. The solar filter was installed in front of the SIS receivers. It attenuated the radio emission by about 4 dB at 100 GHz. In this situation, the difference between the ambient temperature levels and sky level was too small. Hence, the beam chopper method (Penzias and Burrus 1973) could not be used to calibrate the observed signal. In response to this issue, we normalized the observed signal using the quiet region (described below in Section 3.1).

## 3. OBSERVATION AND DATA ANALYSIS
### 3.1 Observation

The observation was carried out on 2014 February 12. Figure 1(a) shows the solar disc image at UV wavelength range (1700 Å), observed by the Atmospheric Imaging Assembly (AIA; Lemen et al. 2012) onboard the Solar Dynamics Observatory (SDO). The radio contour map at 115 GHz, observed by the 45 m telescope, is overlaid on Figure 1(b). The two red lines in Figure 1(b) indicate the right ascension direction. We observed the area between the two red lines using raster-scans along the right ascension direction. Each scan had a length of 3000" and lasted 80 s. Hence, the scan speed was 37.5"/s. The detected signal was sampled by a 10 Hz analog-to-digital converter (3.75"/sample). A map contained 32 scans separated by 7". Therefore, the Nyquist sampling was performed for both right ascension and declination directions. It took 50 min to observe the entire region shown in Figure 1(b). The center of the solar disc rotated 7.8" during the observation, which is about half of the beam size at 115 GHz. We used the AIA image taken at the time when the scan crossed over the center of the sunspot region. Hence, we can neglect the rotation of the disc when comparing the radio and UV images at the sunspot region. The Geostationary Operational Environmental Satellites (GOES) observed no C or larger class flares in this region during the observational period.

The pointing errors of each scan were corrected by using the solar limb of the UV image at 1700 Å as a reference. We co-aligned two edges of the solar disc of the radio scan profiles and the corresponding UV data. Raster-scans that passed through the solar center along the declination direction were inserted every 1 h to correct the pointing errors of the declination. The white lines in Figure 1(b) are the limbs of the co-aligned radio contour map. The actual limb at millimeter range and at 1700 Å may be different. Ewell et al. (1993) suggests that the limb at submillimeter range is 3000–4000 km (about 4–6") above the visible limb. The limb at millimeter range can be slightly higher than the submillimeter limb. However, the beam size of the telescope is about 15" at 115 GHz. Hence, the difference of the limb location between the millimeter range and 1700 Å is smaller than the beam size.

The solar filter prevented the measurement of the absolute brightness temperature. Therefore, we used the quiet Sun level as a reference for the brightness temperature. We defined the quiet region using the UV image at 1700 Å, as indicated by the red rectangle in Figure 1(a), which designates the region that contains neither sunspot nor plage. The typical signal level of each scan changed because of the variation of weather conditions. We assumed the weather conditions to be constant during an individual scan, and the average brightness temperature to be constant throughout the quiet region. We normalized the signal level of each scan using the assumption that the average signal level of the quiet region contained in the individual scan should be constant. We then made a histogram of the signal level in the quiet region (red rectangle in Figure 1(a)). We fitted the histogram according to a Gaussian function. The center of the Gaussian function is defined as the quiet Sun level. The brightness temperatures of the quiet Sun we used were 7500 K at 115 GHz and 7900 K at 85 GHz (Selhorst et al. 2005). The scan should start and end at sky level, where the emission of the Sun should be 0 K. If the start and end of a scan had different levels, then the background was assumed to change linearly, and was subtracted from the scan. Finally, the level of the scan was normalized using the sky level and the level of the quiet Sun.

### 3.2 Sunspot Umbra

Figure 2 shows the close-up image around the active region AR 11976, which is indicated by the red arrow in Figure 1(a). The left and right panels show color contour maps of 85 GHz and 115 GHz, respectively. From these images, it is evident that there is a high correlation between the brightness temperature of the millimeter waves and the UV emission. For example, the plage regions (high UV emission) show high brightness temperature, whereas the quiet regions and the sunspot region (low UV emission) show relatively low brightness temperature. The average brightness temperature of the sunspot umbra region, which is surrounded by the white rectangle (region U) in Figure 2(a), is 7830 K at 85 GHz and 7430 K at 115 GHz.

### 3.3. Fine Structure of the Quiet Region

It should also be mentioned that we found a high correlation between the UV 1700 Å emission and millimeter radio emission at the quiet region. Figure 3 shows the comparison between the UV 1700 Å and 115 GHz emission at a quiet region. To compare the radio and UV data, which have different spatial resolutions, we assumed the main beam of the 45 m telescope to be a Gaussian with a full width half maximum (FWHM) of 15" at 115 GHz, and convolved it to the UV 1700 Å image (Figure 3(a)). Figure 3(c) shows the 1700 Å image convolved with the main beam of the 45 m telescope. The brightness temperature at 115 GHz is overlaid in Figure 3(d). It is clear that there is a close one-to-one correlation between the radio and UV images.

## 4. DISCUSSION
### 4.1. Evaluation of the Weather and Instrument Conditions

We performed the same 32 raster-scan observations twice to evaluate the time variation of the weather and instrument conditions. Figure 4 shows a pair of scan profiles at the same region of the Sun derived from the two raster-scans. These two scans are separated by approximately 1 h. The difference of the brightness temperatures of the two scans at the quiet region is about 1.5%. This difference is thought to be mainly due to the weather and/or instrument conditions since there were no flares along the scanned region during the two observations. This difference of the brightness temperatures is the same level in the other pairs of scans (less than 2.0% in all pairs of scans). Therefore, the error due to the weather and/or instrument conditions is assumed to be less than 2% in this study.

Figure 5 shows the histogram of the brightness temperature of the quiet region in Figure 1(a). In this histogram, the signal level has already been converted to the brightness temperature. We fitted the histograms using Gaussian functions (red curves). The centers of the Gaussian functions represent the quiet Sun levels (7900 K at 85 GHz and 7500 K at 115 GHz). The blue lines indicate the average brightness temperatures of the sunspot umbra region (region U) in Figure 2(a). They are 7830 K at 85 GHz and 7430 K at 115 GHz. Hence, the brightness temperature of the sunspot is 70 K lower than the quiet region at 85 GHz and 115 GHz.

The arrows in Figure 5 indicate the error bars ($\pm 2\%$) of the brightness temperatures of the sunspot umbra. The difference between the brightness temperatures of the sunspot umbra (blue lines) and quiet regions (centers of the Gaussian functions) in this study is about 1%, which is within the specified margin of error. The quiet region also exhibits brightness temperature variation, as shown in Figure 3. The gray regions in Figure 5 indicate the FWHM of the Gaussian functions of the histograms. The brightness temperature of the umbra region is included in the FWHM of the brightness temperature of the quiet region. From these results, it is suggested that the brightness temperature of the sunspot umbra is almost the same as that of the quiet region.

### 4.2. Evaluation of the Side Lobes

As a result of solar eclipse observations, the solar limb is known to have a sharp edge in the millimeter range (Ewell et al. 1993). However, the scan profile observed by the 45 m telescope shows a gradual limb profile in the brightness temperature (see between 900 and 1300 arcsec in Figure 4). Millimeter telescopes usually have broad side lobes (so-called "far-wings") that make the observed limb profile gradually drop to zero (e.g. Bastian et al. 1993). Figure 4 shows that the 45 m telescope also has far-wings. The width of the scan region (vertical to the scan direction) was about 200" in this study. However, Figure 4 shows that the gradual limb profile extends more than 500" outside of the limb. That means the width of the broad far-wings was much larger than the scan width. Hence, it is difficult to two-dimensionally deconvolve the beam pattern from the measured map. For the future studies, a wider scan region will be required. Consequently, this will require faster scan speeds, which will be important for avoiding the change in weather conditions.

The actual brightness temperature of the sunspot umbra should be lower than our observational result. This discrepancy is due to the fact that sunspots are usually surrounded by a brighter plage region, as shown in Figure 2, which allows the side lobes to contaminate the main beam. Hence, the observed brightness temperature of the sunspot should be an upper limit on the true value. The main beam efficiencies of the 45 m telescope in this observation are about 44% at 85 GHz and 29% at 115 GHz.[1] Hence, the contamination of the side lobes at 115 GHz is larger than the contamination at 85 GHz.

### 4.3. Comparison with Observational and Modeling results

Our observational result suggests that the upper limit of the brightness temperature at the sunspot umbra is almost the same as that of the quiet region at 85 GHz (3.5 mm) and 115 GHz (2.6 mm). This result is consistent with the interferometer (BIMA) observation at 3.5 mm (White et al. 2006). It is also suggested that, at the submillimeter wavelength range, the brightness temperature of a sunspot umbra is lower than that of the quiet region from measurements at 0.85 mm using the CSO (Bastian et al 1993), and at 0.35, 0.85, and 1.2 mm using the JCMT (Lindsey & Kopp 1995). Our observational result and previous observational studies consistently derived the result that a sunspot umbra is darker than the quiet region between the submillimeter and millimeter range.

There are many atmospheric models of sunspots (e.g. Avrett 1981; Maltby et al. 1986; Severino et al. 1994; Socas-Navarro 2007; Fontenla et al. 2009). Loukitcheva et al. (2014) calculated the brightness temperature of sunspot umbrae from these models. They suggested that sunspots are darker than the quiet region at submillimeter range, which is consistent with the submillimeter observations (Bastian et al 1993; Lindsey & Kopp 1995). On the other hand, they suggested that the brightness temperature of sunspot umbra is higher than that of the quiet region at millimeter range in all calculated models. This is inconsistent with our observational results at millimeter range.

---

[1] http://www.nro.nao.ac.jp/~nro45mrt/html/prop/eff/eff2013.html

The free-free emission from the Sun at a given wavelength has a broad height distribution of the emission formation region (emission contribution function). At the submillimeter range, the emission originates mainly from the upper photosphere and the temperature minimum region (Wedemeyer-Böhm et al. 2007). Hence, it seems that many atmospheric models can extract the observational results between the photosphere and the temperature minimum region. On the other hand, the emission around the 3 mm range is formed between the chromosphere and the transition region. Therefore, our observational result requires us to improve the current sunspot atmospheric models between the chromosphere and transition region. These atmospheric models are based on the observation of optical and UV lines that formed under non-LTE conditions in the chromosphere. Hence, it is difficult to derive atmospheric information from such observational results, and such models can contain a large amount of uncertainty.

Our observation only contains two wavelengths around 3 mm, which is insufficient for deconvolving an atmospheric model from the observations. However, it is worth mentioning that one possible way of improving the atmospheric models involves a height of the transition region. The calculation of Loukitcheva et al. (2014) shows that the brightness temperature of sunspot umbrae by Severino et al. (1994) is relatively lower than those of the other models that they investigated. Even though the brightness temperature of umbrae in Severino et al. (1994) is almost the same as that of the quiet region, this model is the closest to our observational results. In this model, the height of the transition region is relatively lower than that of the other models. This model suggests a relatively lower brightness temperature of umbrae between 3 and 10 mm range. Multi-wavelength observation from millimeter to centimeter range will be effective to verify the vertical structure models of the upper chromosphere more definitely.

### 4.4. Brightness Temperature Variation of the Quiet Region

We defined the average brightness temperature in the red rectangle in Figure 1(a) as the brightness temperature of the quiet region. However, the brightness temperature varies throughout this region, as shown in Figure 3(d). Figure 3(b) shows the line-of-sight magnetic field observed with the Helioseismic and Magnetic Imager (HMI: Scherrer et al. 2012) onboard SDO. Comparing these two figures, the radio brightness temperature seems to correspond to the network of the magnetic field structure in the quiet region.

The lowest brightness temperature of the quiet region that is shown in the white rectangle in Figure 2(a) (region M) is 7290 K at 115 GHz, which is lower than that of the umbra by 140 K. It should also be noted that the quiet region observation was also affected by the side lobes. Hence, we cannot distinguish the network structure region from the surrounding region. Therefore, we defined a brightness temperature of the quiet region using the histogram of the signal level in the region shown in Figure 1(a). The minimum brightness temperature of the quiet region shown as region M in Figure 2(a) can be an alternative definition of the quiet region in future studies.

### 5. SUMMARY

We observed the various features of the chromosphere at millimeter range using the Nobeyama 45 m telescope at 85 and 115 GHz. The large diameter of the telescope enabled us to observe, for the first time, a sunspot with enough spatial resolution to distinguish the sunspot umbra from the other regions at millimeter range. We used a solar filter that enabled us to prevent saturation of the receivers. However, it prevented us from detecting the actual brightness temperature. Hence, we used the quiet region level as a reference. Our results are summarized as follows:

1. The upper limit of the brightness temperature of the sunspot umbra is almost the same as that of the quiet region. However, the plage region exhibits a higher brightness temperature than the quiet Sun. The 45 m telescope has broad side-lobes, and the sunspot region is surrounded by a brighter plage region. Hence, the actual brightness temperature of the umbra region should be lower than the observational result. This result is inconsistent with the preexisting models, which predict that the sunspot umbra should be brighter than the quiet region at millimeter range.

2. The brightness temperature distribution at millimeter range highly corresponds to the UV 1700 Å emission. At the quiet region, the spatial variation of the radio emission seems to correspond to a network of magnetic field structures.

This study shows that a large single-dish telescope observation is effective for observing the chromosphere at millimeter range. However, large millimeter telescopes usually have broad side-lobes. Hence, precise measurements of broad side-lobes will be essential in deconvolving the actual brightness temperature. It is also important to observe more frequency bands in order to derive accurate brightness temperature spectra.

The observational results of the quiet region in this study predict that there should be a close correlation between the radio brightness temperature at millimeter range and the fine structure of the chromosphere. The Atacama Large Millimeter/submillimeter Array (ALMA) has extremely high spatial resolution at millimeter range (~0.1" at 100 GHz). Moreover, the imaging cadence of ALMA can be as short as 1 s, which is shorter than the Alfven time-scale of the chromosphere. Therefore, this study suggests that ALMA has a significant potential to investigate the various chromospheric phenomena.

## ACKNOWLEDGEMENTS
The 45 m radio telescope is operated by Nobeyama Radio Observatory, a branch of the National Astronomical Observatory of Japan. We wish to thank the 45 m telescope team, including Hiroyuki Iwashita, Tetsuhiro Minamidani, Cheko Miyazawa, Yoshio Tatamitani, and Hiroyuki Nishitani for the support of our observation. SDO data are courtesy of NASA/SDO, as well as the AIA and HMI science teams.

Figures

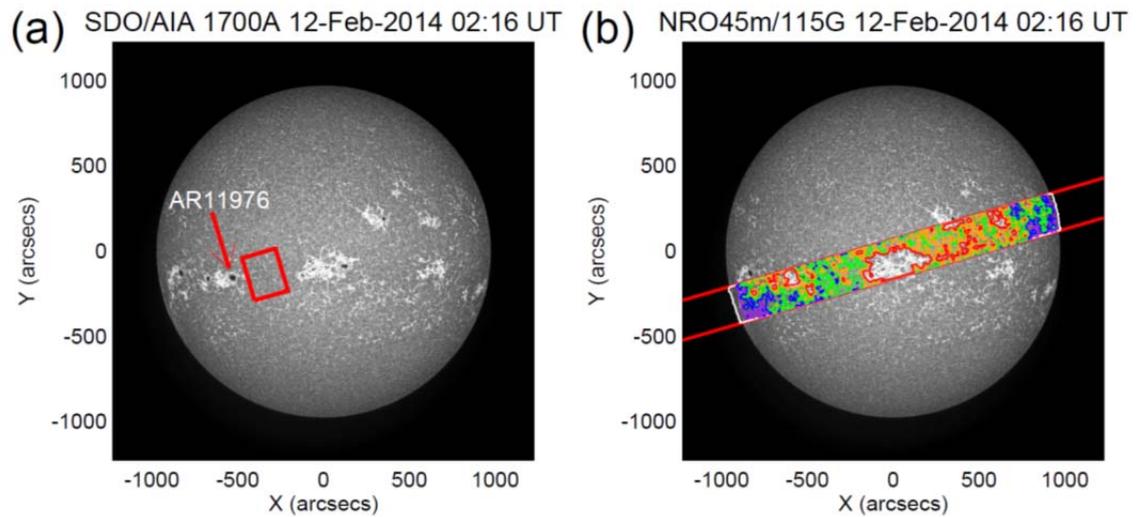

**Figure 1** (a) Full solar disc image at 1700 Å observed on 2014 February 12. (b) Radio contour map observed at 115 GHz overlaid on the 1700 Å image. Red: 7900 K, orange: 7700 K, green: 7500 K, blue: 7300 K, and purple: 7100 K.

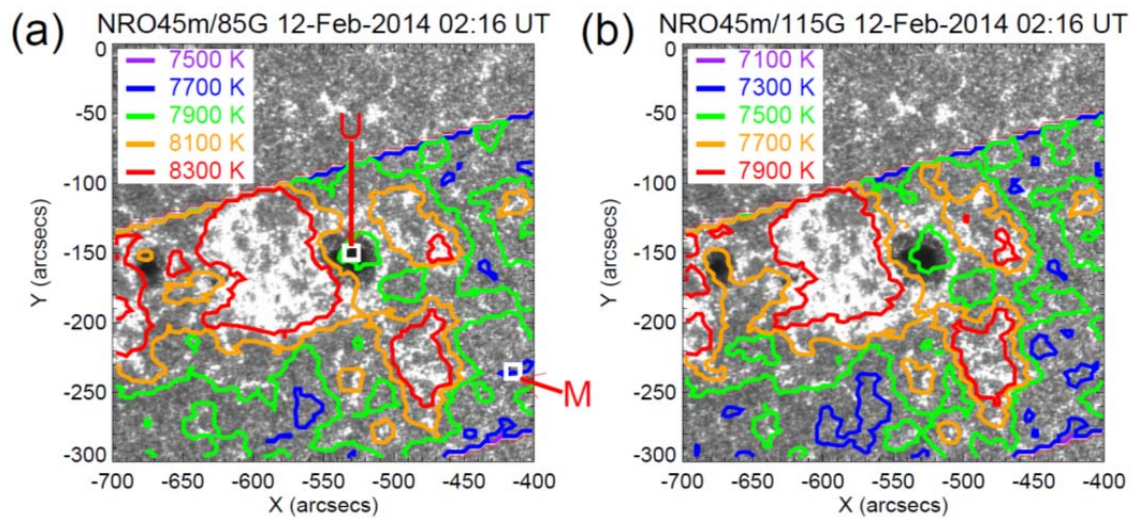

**Figure 2** Radio contour maps around AR 11976, overlaid on 1700 Å images. (a) 85 GHz. (b) 115 GHz.

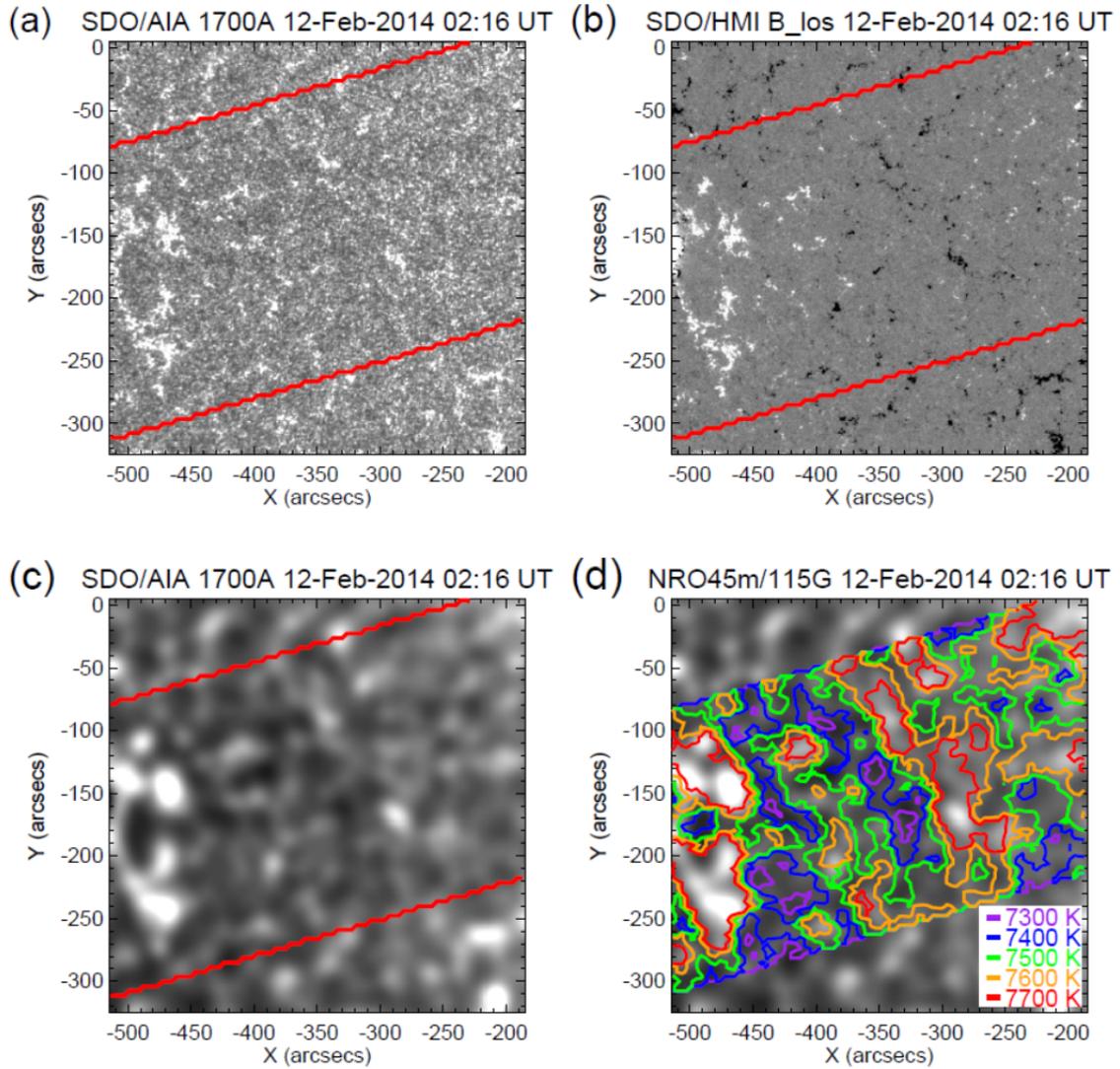

**Figure 3** Comparison between the UV emission, magnetic field, and 115 GHz emission at a quiet region. (a) UV 1700 Å image at the quiet region. (b) Line-of-site magnetic field. (c) UV 1700 Å image that is convolved with the main beam of the 45 m telescope. (d) Brightness temperature at 115 GHz overlaid on Fig. 3(c).

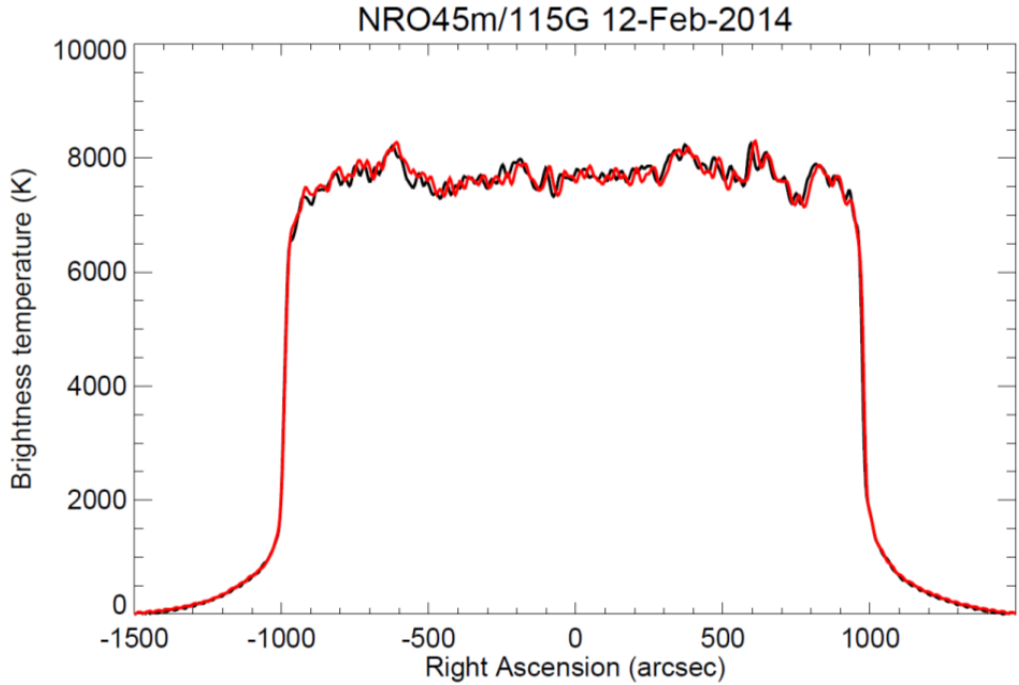

**Figure 4** Two scan profiles along the same region of the Sun at 115 GHz, observed at 01:50 UT (red) and 02:43 UT (black) on 2014 February 12.

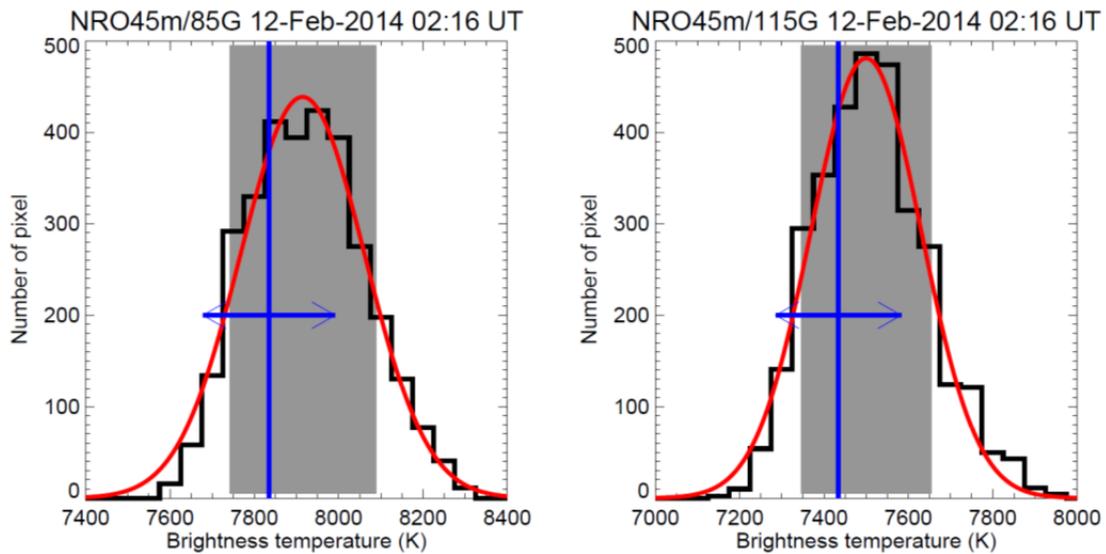

**Figure 5** Histograms of the brightness temperature of the quiet region in Figure 1(a) at (left) 85 GHz and (right) 115 GHz. Gray region: a full width half maximum (FWHM) of the Gaussian function. Blue lines: the average brightness temperatures of the sunspot umbra region (region U). Arrows: error bars of the brightness temperature.